\RequirePackage{ifpdf}
\documentclass[11pt,letterpaper]{article}
\pdfoutput=1
\usepackage{jheppub}
\usepackage{amsmath,amssymb,amsfonts,bm,amscd}
\usepackage{graphicx}
\usepackage{verbatim}
\usepackage{fancybox}
\usepackage{changepage}
\usepackage{slashed}
\usepackage{url}
\usepackage{array}
\usepackage{hhline}
\usepackage{tabularx}
\usepackage{tikz}

\usepackage{float}

\usepackage{rotating,booktabs}
\usepackage[verbose]{placeins}
\usepackage{blindtext}

\graphicspath {{figures/}}

\newcommand{\bea}{\begin{eqnarray}}
\newcommand{\eea}{\end{eqnarray}}
\newcommand{\be}{\begin{equation}}
\newcommand{\ee}{\end{equation}}

\newcommand{\pbra}[1]{\left(#1\right)}
\newcommand{\tq}{\bar q}

\newcommand{\Z}{{\mathbb Z}}
\newcommand{\R}{{\mathbb R}}
\newcommand{\C}{{\mathbb C}}

\newcolumntype{x}[1]{>{\centering\arraybackslash}p{#1}}

 \newcommand*{\matminus}{%
  \leavevmode
  \hphantom{0}%
  \llap{%
    \settowidth{\dimen0 }{$0$}%
    \resizebox{1.1\dimen0 }{\height}{$-$}%
  }%
}

\def\Tr{{\rm Tr \,}}

\def\ch{\check h}

\def\frak{\mathfrak}

\def\tilde{\widetilde}
\def\hat{\widehat}
\def\bar{\overline}

\def\CA{{\mathcal A}}

\def\CH{{\mathcal H}}
\def\CI{{\mathcal I}}

\def\CM{{\mathcal M}}
\def\CN{{\mathcal N}}

\def\CS{{\mathcal S}}
\def\CT{{\mathcal T}}

\def\CV{{\mathcal V}}
\def\CW{{\mathcal W}}

\renewcommand{\bar}{\overline}
\renewcommand{\hat}{\widehat}

\def\^{{\wedge}}
\def\*{{\star}}

\newcommand{\beq}{\begin{equation}}
\newcommand{\eeq}{\end{equation}}

\newcommand{\T}{\mathbb{T}}

\title{A 3d-3d appetizer}
\author{Du Pei and Ke Ye}

\affiliation{Walter Burke Institute for Theoretical Physics, 
California Institute of Technology, \\Pasadena, CA 91125}

\emailAdd{pei@caltech.edu}
\emailAdd{kye@caltech.edu}

\abstract{We test the 3d-3d correspondence for theories that are labeled by Lens spaces. We find a full agreement between the index of the 3d $\CN=2$ ``Lens space theory'' $T[L(p,1)]$ and the partition function of complex Chern-Simons theory on $L(p,1)$. In particular, for $p=1$, we show how the familiar $S^3$ partition function of Chern-Simons theory arises from the index of a free theory. For large $p$, we find that the index of $T[L(p,1)]$ becomes a constant independent of $p$. In addition, we study $T[L(p,1)]$ on the squashed three-sphere $S^3_b$. This enables us to see clearly, at the level of partition function, to what extent $G_\C$ complex Chern-Simons theory can be thought of as two copies of Chern-Simons theory with compact gauge group $G$. 
\\
\\
\\
\\
\\
\\
\\
\\
\\
\\
\\
{\tt CALT-TH-2015-013}}

\begin{document}

\maketitle

\section{Introduction}

The 3d-3d correspondence is an elegant relation between 3-manifolds and three-dimensional field theories \cite{DGH, Terashima:2011qi, Dimofte:2011ju, Dimofte:2011py}. The general spirit is that one can associate a 3-manifold $M_3$ with a 3d $\CN=2$ superconformal field theory $T[M_3; G]$, obtained by compactifying the 6d (2,0) theory on $M_3$
\be\begin{matrix}
\text{6d (2,0) theory on $M_3$}\\
  \text{\rotatebox[origin=c]{-90}{$\leadsto$}} \\
\text{3d $\CN=2$ theory $T[M_3]$}.\end{matrix}
\ee
In this procedure, the 6d theory is topologically twisted along $M_3$ to preserve $\CN=2$ supersymmetry. As a consequence, the 3d $\CN=2$ theory $T[M_3; G]$ only depends on the topology of $M_3$ and the simply-laced Lie algebra $\frak{g}=\mathrm{Lie}G$ that labels the 6d theory\footnote{The theory doesn't depend on small deformations of the metric, but could, in principle, depend on a set of discrete variables, and we already know that a choice of ``framing'' will change $T[M_3]$. In fact, based on current evidence, it is tempting to conjecture that the topology of $M_3$ and the choice of framing completely determine $T[M_3]$. }. Although the dictionary between the dynamics of $T[M_3]$ and topological properties of $M_3$ is incredibly rich \cite{DGH, Dimofte:2011ju, Dimofte:2011py, Beem:2012mb, Dimofte:2013iv, Dimofte:2014zga} and only partially explored, there are two very fundamental relations between $M_3$ and $T[M_3]$. Firstly, the moduli space of supersymmetric vacua of $T[M_3; G]$ on $\R^2\times S^1$ is expected to be homeomorphic to the moduli space of flat $G_\C$-connections on $M_3$:
\be\label{Vacua}
\CM_{\text{SUSY}}(T[M_3;G]) \simeq \CM_{\text{flat}}(M_3;G_\C).
\ee
Second, the partition function of $T[M_3]$ on Lens space $L(k,1)$ should be equal to the partition function of complex Chern-Simons theory on $M_3$ at level $k$ \cite{Cordova:2013cea, Dimofte:2014zga}:
\be\label{Partition}
Z_{T[M_3;G]}[L(k,1)_b]=Z_{\text{CS}}^{(k,\sigma)}[M_3;G_\C].
\ee
The level of complex Chern-Simons theory has a real part $k$ and an ``imaginary part''\footnote{We use the quotation mark here because $\sigma$ can be either purely imaginary or purely real as pointed out in \cite{Witten:1989ip}.} $\sigma$, and $\sigma$ is related to the squashing parameter $b$ of Lens space $L(k,1)_b=S^3_b/\Z_k$ by
\be
\sigma=k\cdot \frac{1-b^2}{1+b^2}.
\ee
For $k=0$, $L(k,1)=S^1\times S^2$, and the equation \eqref{Partition} maps the superconformal index of $T[M_3]$ to partition function of complex Chern-Simons theory at level $(0,\sigma)$ \cite{Dimofte:2011py}
\be\label{Index}
\text{Index}_{T[M_3;G]}(q)=\Tr(-1)^F q^{\frac{E+j_3}{2}}=Z_{CS}^{(0,\sigma)}[M_3;G_\C].
\ee

Despite its beauty and richness, the 3d-3d correspondence has been haunted by many problems since its birth. For example, the theories $T_{\text{DGG}}[M_3]$ originally proposed in \cite{Dimofte:2011ju} miss many branches of flat connections and therefore fail even the most basic test \eqref{Vacua}. This problem was revisited and partially corrected in \cite{Chung:2014qpa}. As for \eqref{Partition} and \eqref{Index}, there is simply no known proposal for $T[M_3]$ associated to \emph{any} $M_3$ that passes these stronger tests. Even the very first non-trivial example of partition function in Chern-Simons theory found in Witten's seminal paper \cite{Witten:1988hf},
\be\label{S3}
Z_{\text{CS}}[S^3;SU(2),k]=\sqrt{\frac{2}{k+2}}\sin\left(\frac{\pi}{k+2}\right),
\ee
has yet to find its home in the world of 3d $\CN=2$ theories.

In \cite{equivariant}, a candidate for the 3d theory $T[L(p,1)]$ was proposed and studied\footnote{More precisely, this is the UV CFT that can flow to numerous different IR theories labelled by UV R-charges of $\Phi$. The IR theory relevant for the 3d-3d relation is given by $R(\Phi)=2$.}:
\be\label{LensTheory}
T[L(p,1);G]\quad {=}\quad \boxed{\genfrac{}{}{0pt}{}{\text{3d $\CN=2$ $G$ super-Chern-Simons theory at level $p$}}{\text{+ adjoint chiral multiplet $\Phi$}}}. 
\ee
This theory was used to produce Verlinde formula, the partition function of Chern-Simons theory on $S^1\times \Sigma$, along with its ``complexification'' --- the ``equivariant Verlinde formula''. Therefore, one may wonder whether this theory could also give the correct partition function of Chern-Simons theory on $S^3$ in \eqref{S3} and its complex analog:
\be\label{S3C}
Z_{\text{CS}}[S^3;SL(2,\C),\tau,\bar{\tau}]=\sqrt{\frac{4}{\tau\bar{\tau}}}\sin\left(\frac{2\pi}{\tau}\right)\sin\left(\frac{2\pi}{\bar{\tau}}\right).
\ee
Here we have used holomorphic and anti-holomorphic coupling constants
\be
\tau=k+\sigma,\quad \bar{\tau}=k-\sigma.
\ee
Indeed, according to the general statement of the 3d-3d correspondence, $T[L(p,1)]$ needs to satisfy 
\be\label{LensLens}
Z_{T[L(p,1);G]}[L(k,1)_b]=Z_{\text{CS}}^{(k,\sigma)}[L(p,1);G_\C]
\ee
and
\be\label{IndexLens}
\text{Index}_{T[L(p,1);G]}(q)=\Tr(-1)^F q^{\frac{E+j_3}{2}}=Z_{CS}^{(0,\sigma)}[L(p,1);G_\C].
\ee
And if we take $p=1$, the above relation states that the index of $T[S^3]$ should give the $S^3$ partition function of complex Chern-Simons theory. Even better, as there is a conjectured duality \cite{Jafferis:2011ns, Kapustin:2011vz} relating this theory to free chiral multiplets, one should be able to obtain \eqref{S3} and \eqref{S3C} by simply computing the index of a free theory! This relation, summarized in diagrammatic form below, 
\be\label{CSFree}
\boxed{\genfrac{}{}{0pt}{}{\text{Chern-Simons}}{  \text{theory on $S^3$}}}
\quad \overset{\text{3d-3d}}{\longleftrightarrow} \quad
\boxed{\genfrac{}{}{0pt}{}{\text{Index of }}{\text{$T[S^3]$}}} \quad \overset{\text{duality}}{\longleftrightarrow}\quad \boxed{\genfrac{}{}{0pt}{}{\text{free chiral}}{\text{multiplets}}}
\ee
will be the subject of section \ref{Sec:S3}. We start section \ref{Sec:S3} by proving the duality (at the level of superconformal index) in \eqref{CSFree} for $G=U(N)$ and then ``rediscover'' the $S^3$ partition function of $U(N)$ Chern-Simons theory from the index of $N$ free chiral multiplets. Then in section \ref{Sec:Ind} we go beyond $p=1$ and study theories $T[L(p,1)]$ with higher $p$. We check that the index of $T[L(p,1)]$ gives precisely the partition function of complex Chern-Simons theory on $L(p,1)$ at level $k=0$. In addition, we discover that index of $T[L(p,1)]$ has some interesting properties. For example, when $p$ is large, 
\be
\mathrm{Index}_{T[L(p,1); U(N)]}=(2N-1)!!
\ee
is a constant that only depends on the choice of the gauge group. In the rest of section \ref{Sec:Ind}, we study $T[L(p,1)]$ on $S^3_b$ and use the 3d-3d correspondence to give predictions for the partition function of complex Chern-Simons theory on $L(p,1)$ at level $k=1$. 

\section{Chern-Simons theory on $S^3$ and free chiral multiplets}\label{Sec:S3}

According to the proposal \eqref{LensTheory}, the theory $T[S^3]$ is $\CN=2$ super-Chern-Simons theory at level $p=1$ with an adjoint chiral multiplet. If one takes the gauge group to be $SU(2)$, this theory was conjectured by Jafferis and Yin to be dual to a free $\CN=2$ chiral multiplet \cite{Jafferis:2011ns}. The Jafferis-Yin duality has been generalized to higher rank groups by Kapustin, Kim and Park \cite{Kapustin:2011vz}. For $G=U(N)$, the statement of the duality is:
\be\label{Duality}
T[S^3]=\boxed{\genfrac{}{}{0pt}{}{\text{$U(N)_1$ super-Chern-Simons theory}}{\text{+ adjoint chiral multiplet}}} \quad \overset{\text{duality}}{\longleftrightarrow}\quad \boxed{\genfrac{}{}{0pt}{}{\text{$N$ free chiral}}{\text{multiplets}}}\,.
\ee
In \cite{equivariant}, a similar duality was discovered\footnote{In \cite{equivariant}, the adjoint chiral is usually assumed to be massive, which introduces an interesting ``equivariant parameter'' $\beta$. Here we are more concerned with the limit where that parameter is zero.}:
\be\label{VortexDuality}
T[L(p,1)]=\boxed{\genfrac{}{}{0pt}{}{\text{$U(N)_p$ super-Chern-Simons theory}}{\text{+ adjoint chiral multiplet}}} \quad \overset{\text{duality}}{\longleftrightarrow}\quad \boxed{\genfrac{}{}{0pt}{}{\text{sigma model to}}{\text{vortex moduli space $\CV_{N,p}$}}}\,.
\ee
Here,  
\be\label{VortexMod}
\CV_{N,p} \; \cong \; \left\{(q,\varphi)\big|\zeta\cdot\mathrm{Id}=q q^{\dagger} +[\varphi,\varphi^\dagger]\right\}/U(N),
\ee
with $q$ being an $N\times p$ matrix, $\varphi$ an $N\times N$ matrix and $\zeta\in\R^+$ the ``size parameter,'' was conjectured to be the moduli space of $N$ vortices in a $U(p)$ gauge theory \cite{Hanany:2003hp}. For $p=1$, it is a well known fact that (see, e.g. \cite{jaffe1980vortices})
\be
\CV_{N,1}\simeq \mathrm{Sym}^N(\C) \simeq \C^N.
\ee
This is already very close to proving that $T[L(1,1); U(N)]=T[S^3; U(N)]$ is dual to $N$ free chirals, with only one missing step. In order to completely specify the sigma model, one also needs to determine the metric on this space. A sigma model to $\C^N$ with the flat metric is indeed a free theory, but it is not obvious that the metric on $\CV_{N,1}$ is flat\footnote{$\CV_{N,p}$ can be obtained using K\"ahler reduction from $\C^{N(N+p)}$ as in \eqref{VortexMod}, and a K\"ahler metric is also inherited in this process. However, this metric on $\CV_{N,p}$ is not protected from quantum corrections. The quantum metric is yet unknown to the best of our knowledge, but for the JY-KKP duality to be true, it should flow to a flat metric in the IR for $p=1$ --- a somewhat surprising prediction.}. However, as the superconformal index of a sigma model only depends on topological properties of the target space, one obtains that 
\be\label{IndIden}
\text{index of $T[S^3; U(N)]$ }=\text{ index of $N$ free chirals},
\ee
proving the duality in \eqref{CSFree} at the level of index. Combining \eqref{IndIden} with the 3d-3d correspondence, one concludes that the index of free chirals equals the $S^3$ partition functions of Chern-Simons theory. This is what we will explicitly demonstrate in this section. 

\subsubsection*{Chern-Simons theory on the three-sphere}

The partition function of $U(N)$ Chern-Simons theory on $S^3$ is 
\be\label{CSS3}
Z_{\text{CS}}\left(S^3;U(N),k\right)=\frac{1}{(k+N)^{N/2}}\prod_{j=1}^{N-1}\left[\sin\frac{\pi j}{k+N}\right]^{N-j}.
\ee
For $N=2$, this gives back \eqref{S3} for $SU(2)$ (modulo a factor coming from the additional $U(1)$). It is convenient to introduce
\be\label{DefQ1}
q=e^{\frac{2\pi i}{k+N}},
\ee
the variable commonly used for the Jones polynomial, and express \eqref{CSS3} as (mostly) a polynomial in $q^{1/2}$ and $q^{-1/2}$:
\be
Z_{\text{CS}}\left(S^3;U(N),k\right)=C\cdot\left(\ln q\right)^{N/2}\prod_{j}^{N-1}\left[q^{j/2}-q^{-j/2}\right]^{N-j}.
\ee
Here $C$ is a normalization factor that does not depend on $q$ and such factors will be dropped in many later expressions without comment.

One can easily obtain the partition function for $GL(N,\C)$ Chern-Simons theory by noticing that it factorizes into two copies of \eqref{CSS3} at level $k_1=\tau/2$ and $k_2=\bar{\tau}/2$ 
\be\label{CCSS3}
Z_{\text{CS}}\left(S^3;GL(N,\C)\right)=\left(\ln q \ln\bar{q}\right)^{N/2}\prod_{j=1}^{N-1}\left[q^{j/2}-q^{-j/2}\right]^{N-j}\left[\bar{q}^{-j/2}-\bar{q}^{j/2}\right]^{N-j}.
\ee
Here, in slightly abusive use of notation (\emph{cf.} \eqref{DefQ1}),
\be
q=e^{\frac{4\pi i}{\tau}},\quad \bar{q}=e^{\frac{4\pi i}{\bar{\tau}}}.
\ee
Notice that the quantum shift of the level $k\rightarrow k+N$ in $U(N)$ Chern-Simons theory is absent in the complex theory \cite{Witten:1989ip, bar-natan1991, Gukov:2008ve}. Although \eqref{CCSS3} is almost a polynomial, it contains ``$\ln q$'' factors. So, at this stage, it is still somewhat mysterious how \eqref{CCSS3} can be obtained as the index of any supersymmetric field theory.

In \eqref{CCSS3} the level is arbitrary and the $k=0$ case is naturally related to superconformal index of $T[S^3]$ \eqref{IndexLens}. For $k=0$,
\be
q=e^{\frac{4\pi i}{\sigma}},\quad \bar{q}=e^{-\frac{4\pi i}{\sigma}}=q^{-1},
\ee
and
\be\label{CCSS30}
Z_{\text{CS}}^{(0,\sigma)}\left(S^3;GL(N,\C)\right)=\left(\ln q\right)^{N}\prod_{j=1}^{N-1}\left[(1-q^{j})(1-q^{-j})\right]^{N-j}.
\ee
This is the very expression that we want to reproduce from the index of free chiral multiplets. 

\subsubsection*{Index of a free theory}

The superconformal index of a 3d $\CN=2$ free chiral multiplet only receives contributions from the scalar component $X$, the fermionic component $\bar{\psi}$ and their $\partial_{+}$ derivatives. If we assume the R-charge of $X$ to be $r$, then the R-charge of $\bar{\psi}$ is $1-r$ and the superconformal index of this free chiral is given by
\be\label{FreeChiral}
\CI_{r}(q)=\prod_{j=0}^\infty\frac{1-q^{1-r/2+j}}{1-q^{r/2+j}}.
\ee
In the $j$-th factor of the expression above, the numerator comes from fermionic field $\partial^j \bar{\psi}$ while the denominator comes from bosonic field $\partial^j X$. Here $q$ is a fugacity variable that counts the charge under $\frac{E+j_3}{2}=R/2+j_3$ and it is the expectation of the 3d-3d correspondence \cite{Dimofte:2011py} that this $q$ is mapped to the ``$q$'' in \eqref{CCSS30}, which justifies our usage of the same notation for two seemingly different variables. Now the only remaining problem is to decide what are the R-charges for the $N$ free chiral multiplets. 

The UV description of theory $T[L(p,1)]$ has an adjoint chiral multiplet $\Phi$ and in general one has the freedom of choosing the R-charge of $\Phi$. Different choices give different IR fix points which form an interesting family of theories. As was argued in \cite{equivariant} using brane construction, the natural choice --- namely the choice that one should use for the 3d-3d correspondence --- is $R(\Phi)=2$. For example, in order to obtain the Verlinde formula, it is necessary to choose $R(\Phi)=2$ while other choices give closely related yet different formulae. As the $N$ free chirals in the dual of $T[S^3;U(N)]$ are directly related to $\Tr\Phi$, $\Tr\Phi^2$, \ldots, $\Tr\Phi^N$, the choice of their R-charges should be 
\be\label{FreeRCharge}
r_m=R(X_m)=2m, \text{ for $m=1,2,\ldots,N$}.
\ee
The index for this assignment of R-charges --- out of the unitarity bound --- contains negative powers of $q$. However, this is not a problem at all because the UV R-charges are mixed with the $U(N)$ flavor symmetries, and $q$ counts a combination of R- and flavor charges. 

One interesting property of the index of a free chiral multiplet \eqref{FreeChiral} is that it will vanish due to the numerator of the $(m-1)$-th factor:
\be
1-q^{m-r_m/2}=0.
\ee
However, there is a very natural way of regularizing it and obtaining a finite result. Namely, we multiply the $q$-independent normalization coefficient $(r_m/2-m)^{-1}$ to the whole expression and turn the vanishing term above into
\be
\lim_{r_m\rightarrow 2m}\frac{1-q^{m-r_m/2}}{r_m/2-m}=\ln q.
\ee
And this is exactly how the ``$\ln q$'' factors on the Chern-Simons theory side arise. With this regularization
\be
\CI_{2m}(q)=\ln q\prod_{j=1}^{m-1}\left[\left(1-q^{-j}\right)\left(1-q^{j}\right)\right],
\ee
and the $2m-1$ factors come from the fermionic fields $\bar{\psi}_m$, $\partial\bar{\psi}_m$,\ldots, $\partial^{2m-2}\bar{\psi}_m$. The contribution of $\partial^{2m-1+l}\bar{\psi}_m$ will cancel with the bosonic field $\partial^{l} X$ as they have the same quantum number. The special log term comes from the field $\partial^{m-1}\bar{\psi}_m$, which has exactly $R+2j_3=0$.

Then it is obvious that
\be
\text{Index}_{T[S^3; U(N)]}=\prod_{m=1}^{N}\CI_{2m}(q)=\left(\ln q\right)^{N}\prod_{j=1}^{N-1}\left[(1-q^{j})(1-q^{-j})\right]^{N-j}
\ee
is exactly the partition function of complex Chern-Simons theory on $S^3$ \eqref{CCSS30}. For example, if $N=1$,
\be
\text{Index}_{T[S^3; U(1)]}=\CI_{2}(q)=\ln q.
\ee
For $N=2$,
\be
\text{Index}_{T[S^3; U(2)]}=\CI_{2}(q)\cdot \CI_{4}(q)=\left(\ln q\right)^2 (1-q^{-1})(1-q).
\ee
To get the renowned $S^3$ partition function of the $SU(2)$ Chern-Simons theory, we just need to divide the $N=2$ index by the $N=1$ index and take the square root:
\be\label{SU2Index}
\sqrt{\frac{\text{Index}_{T[S^3; U(2)]}}{\text{Index}_{T[S^3; U(1)]}}}=\sqrt{\CI_{4}(q)}=-i\cdot \left(\ln q\right)^{1/2} \left(q^{1/2}-q^{-1/2}\right).
\ee
For compact gauge group $SU(2)$, we substitute in  
\be\label{QSU2}
q=e^{\frac{2\pi i}{k+2}}
\ee
and up to an unimportant normalization factor, \eqref{SU2Index} is exactly
\be
Z_{\text{CS}}(S^3; SU(2),k)=\sqrt{\frac{2}{k+2}}\sin\frac{\pi}{k+2}.
\ee

As almost anything in a free theory can be easily computed, one can go beyond index and check the following relation
\be
Z_{\text{$N$ free chirals}}(L(k,1)_b)=Z_{CS}^{(k,\sigma)}(S^3; U(N)).
\ee 
The left-hand side can be expressed as a product of double sine functions \cite{Imamura:2013qxa} and with the right choice of R-charges it becomes exactly the right-hand side, given by \eqref{CSS3}. As this computation is almost identical for what we did with index, we omit it here to avoid repetition.  

Before ending this section, we comment on deforming the relation \eqref{CSFree}. In the formulation of $T[L(p,1)]$ in \eqref{LensTheory}, there is a manifest $U(1)$ flavor symmetry that can be weakly gauged to give an ``equivariant parameter'' $\beta$. And the partition function of $T[L(p,1); \beta]$ should be related to $\beta$-deformed complex Chern-Simons theory studied in \cite{equivariant}:
\be\label{TMatch}
Z_{T[L(p,1);\beta]}(L(k,1))=Z_{\text{$\beta$-CS}}(L(p,1);k).
\ee
When $p=1$, if the JY-KKP duality is true, this $U(1)$ flavor symmetry is expected to be enhanced to a $U(N)$ flavor symmetry of $T[S^3;U(N)]$ that is only visible in the dual description with $N$ free chiral multiplets. Then one can deform $T[S^3]$ by adding $N$ equivariant parameters $\beta_1, \beta_2, \ldots, \beta_N$. It is interesting to ask whether the Chern-Simons theory on $S^3$ naturally admits such an $N$-parameter deformation and whether one can have a more general matching. 
\be\label{GeneralMatch}
\mathrm{Index}_{T[S^3]}(q;\beta_1,\beta_2,\ldots,\beta_N)=Z_{\text{CS}}(S^3; q, \beta_1,\beta_2,\ldots,\beta_N).
\ee
As Chern-Simons theory on $S^3$ is dual to closed string on the resolved conifold \cite{Witten:1992fb, Dijkgraaf:2002fc}, it would also be interesting to understand whether similar deformation of the closed string amplitudes $F_g$ exists.

\section{3d-3d correspondence for Lens spaces}\label{Sec:Ind}

In the previous section, we focused on $T[S^3]$ and found that it fits perfectly inside the 3d-3d correspondence. This theory is the special $p=1$ limit of a general class \eqref{LensTheory} of theories $T[L(p,1)]$ proposed in \cite{equivariant}. In this section, we will test this proposal and see whether it stands well with various predictions of the 3d-3d correspondence. There are several tests to run on the proposed Lens space theories \eqref{LensTheory}. The most basic one is the correspondence between moduli spaces \eqref{Vacua} that one can formulate classically without doing a path integral:
\be\label{SpaceTest}
\CM_{\text{SUSY}}\left(T[L(p,1);U(N)]\right)\simeq \CM_{\text{flat}}\left(L(p,1);GL(N,\C)\right) .
\ee
And our first task in this section is to verify that this is indeed an equality.

\subsection{\texorpdfstring{$\CM_{\text{SUSY}}$ vs. $\CM_{\text{flat}}$}{M\_SUSY vs. M\_flat}} \label{SSec:Space}
 
The moduli space of flat $H$-connections on a three manifold $M_3$ can be identified with the character variety: 
\be
\CM_{\text{flat}}\left(M_3;H\right)\simeq \mathrm{Hom}(\pi_1(M_3),H)/H.
\ee
As $\pi_1(L(p,1))=\Z_p$, this character variety is particularly simple. For example, if we take $H=U(N)$ or $H=GL(N,\C)$ --- the choice between $U(N)$ or $GL(N,\C)$ does not even matter --- this space is a collection of points labelled by Young tableaux with size smaller than $N\times p$. This is in perfect harmony with the other side of the 3d-3d relation where the supersymmetric vacua of $T[L(p,1);U(N)]$ on $S^1\times \R^2$ are also labelled by Young tableaux with the same constraint \cite{equivariant}. We will now make this matching more explicit. 

If we take the holonomy along the $S^1$ Hopf fiber of $L(p,1)$ to be $A$, then
\be
\CM_{\text{flat}}\left(L(p,1);GL(N,\C)\right)\simeq \{A\in GL(N,\C)|A^p=\mathrm{Id}\}/GL(N,\C).
\ee
First we can use the $GL(N,\C)$ action to cast $A$ into Jordan normal form. But in order to satisfy $A^k=\mathrm{Id}$, $A$ has to be diagonal, and each of its diagonal entries $a_l$ has to be one of the $p$-th roots of unity:
\be\label{Holonomy}
a_l^p=1,\text{ for all $l=1,2,\ldots,N$.}
\ee
One can readily identify this set of equations with the $t\rightarrow 1$ limit of the Bethe ansatz equations that determine the supersymmetric vacua of $T[L(p,1);U(N)]$ on $S^1\times \R^2$ \cite{equivariant}:
\be\label{Bethe}
e^{2\pi i p\sigma_l}\prod_{m\neq l}\left(\frac{e^{2\pi i \sigma_l}-te^{2\pi i \sigma_m}}{te^{2\pi i \sigma_l}-e^{2\pi i \sigma_m}}\right)=1,\quad \text{for all of $l=1,2,\ldots,N$}.
\ee
For $t=1$, this equation is simply
\be
e^{2\pi i p\sigma_l}=1, \text{ for $l=1,2,\ldots,N$.}
\ee
And this is exactly \eqref{Holonomy} if one makes the following identification
\be
a_l=e^{2\pi i \sigma_l}.
\ee
Of course this relation between $a_l$ and $\sigma_l$ is more than just a convenient choice. It can be derived using the brane construction of $T[L(p,1)]$. In fact, it just comes from the familiar relation in string theory between holonomy along a circle and positions of D-branes after T-duality. Indeed, in the above expression, the $a_l$'s on the left-hand side label the $U(N)$-holonomy along the Hopf fiber, while the $\sigma_l$'s on the right-hand side are coordinates on the Coulomb branch of $T[L(p,1)]$ after reduction to 2d, which exactly correspond to positions of $N$ D2-branes.

\subsubsection*{$G_\C$ Chern-Simons theory from $G$ Chern-Simons theory}

The fact that $\CM_{\text{flat}}$ is a collection of points is important for us to compute the partition function of complex Chern-Simons theory. Although there have been many works on complex Chern-Simons theory and its partition functions, starting from \cite{Witten:1988hc, Witten:1989ip} to perturbative invariant in \cite{bar-natan1991, Apol}, state integral models in \cite{2007JGP....57.1895H, Dimofte:2009yn, Dimofte:2014zga} and mathematically rigorous treatment in \cite{2011arXiv1109.6295E, 2013arXiv1305.4291E, 2014arXiv1409.1208E}, what usually appear are certain subsectors of complex Chern-Simons theory, obtained from some consistent truncation of the full theory. In general, the {\it full} partition function of complex Chern-Simons theory is difficult to obtain, and requires proper normalization to make sense of. Some progress has been made toward understanding the full theory on Seifert manifolds in \cite{equivariant} using topologically twisted supersymmetric theories. However, if $\CM_{\text{flat}}(M_3; G_\C)$ is discrete and happens to be the same as $\CM_{\text{flat}}(M_3; G)$, then one can attempt to construct the full partition function of the $G_\C$ Chern-Simons theory on $M_3$ from the $G$ Chern-Simons theory. The procedure is the following. One first writes the partition function of the $G$ Chern-Simons theory as a sum over flat connections:
\be
Z^{\text{full}}=\sum_{\alpha\in \CM} Z_{\alpha}.
\ee
And because the action of the $G_\C$ Chern-Simons theory 
\begin{equation}
\begin{aligned}
S = & \frac{\tau}{8\pi} \int {\rm Tr} \left({\cal A} \wedge d{\cal A} + \frac{2}{3} {\cal A} \wedge {\cal A} \wedge {\cal A} \right)\\
+ & \frac{\bar{\tau}}{8\pi} \int {\rm Tr} \left(\bar{\cal  A} \wedge d\bar{\cal A} + \frac{2}{3} \bar{\cal A} \wedge \bar{\cal A} \wedge \bar{\cal A} \right)
\end{aligned}
\end{equation}
is simply two copies of the $G$ Chern-Simons theory action at level $k_1=\tau/2$ and $k_2=\bar{\tau}/2$, one  would have
\be\label{TreePert}
Z_{\alpha}(G_\C;\tau,\bar{\tau})=Z_{\alpha}\left(G;\frac{\tau}{2}\right)Z_{\alpha}\left(G;\frac{\bar{\tau}}{2}\right),
\ee
if $\CA$ and $\bar{\CA}$ were independent fields. So, one would naively expect
\be\label{FullTree}
Z^{\text{full}}(G_\C;\tau,\bar{\tau})=\sum_{\alpha\in \CM} Z_{\alpha}\left(G;\frac{\tau}{2}\right)Z_{\alpha}\left(G;\frac{\bar{\tau}}{2}\right).
\ee
But as $\CA$ and $\bar{\CA}$ are not truly independent, \eqref{FullTree} is in general incorrect and one needs to modify it in a number of ways. For example, as mentioned before, the quantum shift of the level $\tau$ and $\bar{\tau}$ in $G_\C$ Chern-Simons theory is zero, so for $Z_{\alpha}(G)$ on the right-hand side, one needs to at least remove the quantum shift $k\rightarrow k+{\check h}$ in $G$ Chern-Simons theory, where ${\check h}$ is the dual Coxeter number of $\frak{g}$. There may be other effects that lead to relative coefficients between contributions from different flat connections $\alpha$ and the best one could hope for is
\be\label{FullTree1}
Z^{\text{full}}(G_\C;\tau,\bar{\tau})=\sum_{\alpha\in \CM} e^{iC_\alpha} Z'_{\alpha}\left(G;\frac{\tau}{2}\right)Z'_{\alpha}\left(G;\frac{\bar{\tau}}{2}\right),
\ee
where
\be
Z'_{\alpha}\left(G;\frac{\tau}{2}\right)=Z_{\alpha}\left(G;\frac{\tau}{2}-{\check h}\right).
\ee
One way to see that \eqref{FullTree} is very tenuous, even after taking care of the level shift, is by noticing that the left-hand side and the right-hand side behave differently under a change of framing. If the framing of the three-manifold is changed by $s$ units, the left-hand side will pick up a phase factor
\be\label{Framing}
\exp\left[\varphi_\C^{\text{fr.}}\cdot s\right]=\exp\left[ \frac{\pi i (c_L-c_R)}{12}\cdot s \right].
\ee
Here $c_L$ and $c_R$ are the left- and right-moving central charges of the hypothetical conformal field theory that lives on the boundary of the complex Chern-Simons theory \cite{Witten:1989ip}:
\be\label{Central}
(c_L, c_R) = \dim G \cdot  \left(1 - \frac{2 \check h}{\tau}, 1+ \frac{2 \check h}{\bar \tau} \right).
\ee 
The right-hand side of \eqref{FullTree} consists of two copies of the Chern-Simons theory with compact gauge group $G$, so the phase from change of framing is 
\be
\exp\left[\varphi^{\text{fr.}}\cdot s\right]=\exp\left[ \frac{\pi i}{12}\left(\frac{\tau/2-\ch}{\tau/2}+\frac{\bar{\tau}/2-\ch}{\bar{\tau}/2}\right)\dim G\cdot s \right].
\ee
The two phases are in general different
\be
\varphi_\C^{\text{fr.}}-\varphi^{\text{fr.}}=\frac{2\pi i \dim G}{12}.
\ee
So \eqref{FullTree} has no chance of being correct at all and the minimal way of improving it is to add the phases, $C_\alpha$, as in \eqref{FullTree1}, which also transform under change of framing.

It may appear that the expression \eqref{FullTree1} is not useful unless one can find the values of the $C_\alpha$'s. However, as it turns out, for $k=0$ (or equivalently $\tau=-\bar{\tau}$), all of the $C_\alpha$'s are constant, and \eqref{FullTree1} without the $C_\alpha$'s gives the correct partition function\footnote{``Correct'' in the sense that it matches the index of $T[L(p,1)]$. }. This may be closely related to the fact that for $k=0$,
\be
c_L-c_R=-2\check h\dim G\left(\frac{1}{\tau}+\frac{1}{\bar{\tau}}\right)=0.
\ee

\subsection{Superconformal index}

We have shown that the proposal \eqref{LensTheory} for $T[L(p,1)]$ gives the right supersymmetric vacua and we shall now move to the quantum level and check the relation between the partition functions:
\be
\text{Index}_{T[L(p,1);U(N)]}(q)=Z_{CS}\left(L(p,1);GL(N,\C),q\right).
\ee   
We have already verified this for $p=1$ in the previous section. Now we consider the more general case with $p\geq 1$.  

The superconformal index of a 3d $\CN=2$ SCFT is given by \cite{Bhattacharya:2008zy}
\be\label{IndexDef}
\CI(q, t_i) = {\rm Tr} \left[(-1)^F e^{-\gamma (E - R -j_3)} q^{\frac{E+j_3}{2}}  t^{f_i} \right].
\ee
Here, the trace is taken over the Hilbert space of the theory on $\R\times S^2$. Because of supersymmetry, only BPS states with
\be
E-R-j_3=0
\ee
will contribute. As a consequence, the index is independent of $\gamma$ and only depends on $q$ and the flavor fugacities, $t_i$. For $T[L(p,1)]$, there is always a $U(1)$ flavor symmetry and we can introduce at least one parameter $t$. When this parameter is turned on, on the other side of the 3d-3d correspondence, complex Chern-Simons theory will become the ``deformed complex Chern-Simons theory''. This deformed version of Chern-Simons theory was studied on geometry $\Sigma \times S^1$ in \cite{equivariant} and will be studied on more general Seifert manifolds in \cite{fractional}. However, because in this paper our goal is to \emph{test} the 3d-3d relation (as opposed to using it to study the deformed Chern-Simons theory), we will usually turn off this parameter by setting $t=1$, and compare the index $\CI(q)$ with the partition function of the \emph{undeformed} Chern-Simons theory, which is only a function of $q$, as in \eqref{CCSS30}. 

Viewing the index as the partition function on $S^1\times_q S^2$ and using localization, \eqref{IndexDef} can be expressed as an integral over the Cartan $\T$ of the gauge group $G$ \cite{Imamura:2011su}:
\begin{equation}
{\cal I} = \frac{1}{\left| {\cal W} \right|}\sum_{m}   \int \prod_j \frac{dz_j}{2\pi i z_j} e^{-S_{\rm{CS}}(m)} q^{\epsilon_0/2}e^{ib_0(h)} t^{f_0} \exp{ \left[ \sum_{n=1}^{+\infty} \frac{1}{n} {\rm{ Ind}}(z_j^n, m_j; t^n, q^n) \right]}.
\end{equation}
Here $h, m\in \frak{t}$ are valued in the Cartan subalgebra. Physically, $e^{ih}$ is the holonomy along $S^1$ and is parametrized by $z_i$, which are coordinates on $\T$. 
\be
m=\frac{i}{2\pi}\int_{S^2} F
\ee
is the monopole number on $S^2$ and takes value in the weight lattice of the Langlands dual group $^LG$. $\left| {\cal W} \right|$ is the order of the Weyl group and the other quantities are:
\begin{equation}
\begin{aligned}
 b_0(h) & = -\frac{1}{2}\sum_{\rho \in \mathfrak{R}_{\Phi}} \left|\rho(m)\right| \rho(h),\\
 f_0 & = -\frac{1}{2}\sum_{\rho \in \mathfrak{R}_{\Phi}} \left|\rho(m)\right| f,\\
 \epsilon_0 & = \frac{1}{2}\sum_{\rho \in \mathfrak{R}_{\Phi}} (1-r)\left|\rho(m)\right| - \frac{1}{2}\sum_{\alpha \in {\rm ad}(G)} \left| \alpha(m) \right|,\\
 S_{\rm CS} & =  i p\, \mathrm{tr} (m h),
\end{aligned}
\end{equation}
and
\begin{equation}
\begin{aligned}
{\rm Ind}(e^{ih_j}=z_j, m_j; t; q) = & -\sum_{\alpha \in {\rm ad}(G)} e^{i \alpha(h)} q^{\left|  \alpha (m) \right|}\\
                                                    & + \sum_{\rho \in \mathfrak{R}_{\Phi}}\left[ e^{i \rho (h)} t\, \frac{q^{\left|\rho(m)\right|/2 +r/2}}{1-q} - e^{-i \rho (h)} t^{-1} \frac{q^{\left|\rho(m)\right|/2 +1-r/2}}{1-q} \right]
\end{aligned}
\end{equation}
is the ``single particle" index. $\frak{R}_\Phi$ is the gauge group representation for all matter fields. Using this general expression, the index of $T[L(p,1);U(N)]$ can be expressed in the following form:
\begin{equation}\label{IndexInt}
\begin{aligned}
{\cal I}(q,t) = \sum_{{m_1 \geqslant \cdots \geqslant m_N} \in \mathbb{Z}}  \frac{1}{\left| {\cal W}_m \right|} \int \prod_j \frac{dz_j}{2\pi i z_j} & \prod_{i}^{N} \left( z_i \right)^{2 { p}m_i} \prod_{i \neq j}^{N} t^{- \left| m_i -m_j \right|/2} q^{-R \left| m_i -m_j \right|/4} \left( 1-q^{ \left| m_i -m_j \right|/2} \frac{z_i}{z_j} \right) \\
& \prod_{i \neq j}^{N} \frac{\left( \frac{z_j}{z_i} t^{-1} q^{ \left| m_i -m_j \right|/2+1-R/2}; q \right)_{\infty}}{\left( \frac{z_i}{z_j} t q^{ \left| m_i -m_j \right|/2+R/2}; q \right)_{\infty}} \times \left[\frac{(t^{-1}q^{1-R/2}; q)_{\infty}}{(t q^{R/2}; q)_{\infty}}\right]^N.
\end{aligned}
\end{equation}
Here we used the $q$-Pochhammer symbol $(z; q)_n = \prod_{j=0}^{n-1} (1-z q^j)$. ${\cal W}_m\subset {\cal W}$ is the stabilizer subgroup of the Weyl group that fixes $m \in \frak{t}$ and $R$ stands for the R-charge of the adjoint chiral multiplet and will be set to $R=2$ --- the choice that gives the correct IR theory. 

In the previous section, we have found the index for $T[S^3]$ to be exactly equal to the $S^3$ partition function of Chern-Simons theory. There, we used an entirely different method by working with the dual description of $T[L(p,1); U(N)]$, which is a sigma model to the vortex moduli space $\CV_{N,p}$. For $p=1$, this moduli space is topologically $\C^N$ and the index of the sigma model is just that of a free theory. For $p\geq 2$, such a simplification will not occur and the index of the sigma model is much harder to compute\footnote{In general, it can be written as an integral of a characteristic class over $\CV_{N,p}$ that one can evaluate using the Atiyah-Bott localization formula. Similar computations were done in two dimensions in, {\it e.g.} \cite{DGH} and \cite{Bonelli:2011fq}.}. In contrast, the integral expression \eqref{IndexInt} is {\it easier} to compute with larger $p$ than with $p=1$, because fewer topological sectors labelled by the monopole number $m$ contribute. As we will see later, when $p$ is sufficiently large, only the sector $m=(0,0,\ldots,0)$ gives non-vanishing contribution. So the two approaches of computing the index have their individual strengths and are complementary to each other.

Now, one can readily compute the index for any $T[L(p,1); G]$ and then compare ${\cal I}(q,t=1)$ with the partition function of the complex Chern-Simons theory on $L(p,1)$. We will first do a simple example with $G=SU(2)$, to illustrate some general features of the index computation. 

\subsubsection*{Index of $T[L(p,1); SU(2)]$}

We will start with $p=1$ and see how the answer from section \ref{Sec:S3} arises from the integral expression \eqref{IndexInt}. In this case, \eqref{IndexInt} becomes
\begin{equation}\label{SU2Int}
\begin{aligned}
{\cal I} & = \sum_{m \in \mathbb{Z}} \int \frac{dz}{4\pi i z} e^{i h m} q^{-2 |m|} \left( 1-q^{|m|} e^{ih}\right)^2 \left( 1-q^{ |m|} e^{-ih}\right)^2 \prod_{k=0}^{+\infty} \frac{1-q^{k+1-R/2}}{1-q^{k+R/2}}\\
& = \sum_{m \in \mathbb{Z}} \int \frac{dz}{4\pi i z} z^{m} q^{-2 |m| } \left(1+q^{2|m|} - z q^{|m|} - z^{-1} q^{|m|}\right)^2 \left[(R-2) \ln q\right]\\
& = \sum_{m \in \mathbb{Z}} \int \frac{dz}{4\pi i z} z^{m} \left(q^{2|m|} + q^{-2|m|} + 4 - 2\left(z+\frac{1}{z}\right) \left(q^{|m|}+\frac{1}{q^{|m|}}\right) + \left( z^2 + \frac{1}{z^2} \right) \right)\\
&\ \ \ \ \ \ \ \ \ \ \ \ \times  \left[(R/2-1) \ln q\right].
\end{aligned}
\end{equation}
As in section \ref{Sec:S3}, the index will be zero if we naively take $R=2$ because of the $1 - q^{1-r/2}$ factor in the infinite product. When $R\rightarrow 2$, the zero factor becomes
\begin{equation}
1 - q^{1-R/2} = 1 - \exp \left[(1-R/2) \ln q\right] \approx (R/2-1) \ln q.
\end{equation}
As in section \ref{Sec:S3}, we can introduce a normalization factor $(R/2-1)^{-1}$ in the index to cancel the zero, making the index expression finite. 

The integral in \eqref{SU2Int} is very easy to do and the index receives contributions from three different monopole number sectors
\begin{equation}
{\cal I} = \frac{1}{2} \ln q\ ({\cal I}_{m=0} + {\cal I}_{m=\pm 1} + {\cal I}_{m=\pm 2}),
\end{equation}
with
\bea
{\cal I}_{m=0}     &=&  \int \frac{dz}{2\pi i z} \left( q^{0}+q^{-0}+4 \right)=6,\\
{\cal I}_{m=\pm 1} &=&  -2  \sum_{m =\pm 1} \int \frac{dz}{2\pi i z} z^{m} \left(q^{ \left| m \right|}+q^{- \left| m \right|} \right) \left(z+\frac{1}{z} \right)=-4 (q+q^{-1}),
\eea
and
\be
{\cal I}_{m=\pm 2} =  \sum_{m=\pm 2} \int \frac{dz}{2\pi i z} z^{m} \left( z^2 +\frac{1}{z^2} \right)=2.
\ee
So the index is 
\begin{equation}\label{SL2Index}
\begin{aligned}
{\cal I} & =  \frac{1}{2} \ln q \left( 6 - 4(q+q^{-1}) +2 \right)\\
& = - 2\ln q \left( q^{1/2} - q^{-1/2} \right)^2.
\end{aligned}
\end{equation}
Modulo a normalization constant, this is in perfect agreement with results in section \ref{Sec:S3}. Indeed, the square root of \eqref{SL2Index} is identical to \eqref{SU2Index} and reproduces the $S^3$ partition function of the $SU(2)$ Chern-Simons theory,
\be
Z_{CS}(S^3; SU(2),k)=\sqrt{\frac{2}{k+2}}\sin\frac{\pi}{k+2},
\ee
once we set 
\be
q=e^{\frac{2\pi i}{k+2}}.
\ee

It is very easy to generalize the result \eqref{SL2Index} to arbitrary $p$. For general $p$, the index is given by
\begin{equation}
\begin{aligned}
{\cal I} & = \frac{1}{2}\ln q \sum_{m \in \mathbb{Z}} \int \frac{dz}{2\pi i z} z^{ p m}\\
&\ \ \ \times \left( q^{2 \left| m \right|}+q^{-2 \left| m \right|}+4 - 2 \left(q^{ \left| m \right|}+q^{- \left| m \right|} \right) \left( z +\frac{1}{z} \right)+\left( z^2+\frac{1}{z^2} \right) \right).
\end{aligned}
\end{equation}
The only effect of $p$ is to select monopole numbers that contribute. For example, if $p=2$, only $m=0$ and $m=\pm 1$ contribute to the index and we have
\be\label{SU2Limit0}
\CI^{p=2}=\frac{1}{2} \ln q\ ({\cal I}_{m=0} + {\cal I}_{ m=\pm 1}^{p=2})=\frac{1}{2} \ln q\ (6 + 2)=4\ln q.
\ee
If $p>2$, only the trivial sector is selected, and
\be\label{SU2Limit}
\CI(p>2)=\frac{1}{2} \ln q\ {\cal I}_{m=0} =3\ln q.
\ee
This is a general feature of indices of the ``Lens space theory'' and we will soon encounter this phenomenon with higher rank gauge groups.

\subsubsection*{The test for 3d-3d correspondence}

We list the index of $T[L(p,1); U(N)]$, obtained using \textit{Mathematica}, in table \ref{IndexTab}. Due to limitation of space and computational power, it contains results up to $N=5$ and $p=6$. The omnipresent $(\ln q)^N$ factors are dropped to avoid clutter, and after this every entry in table \ref{IndexTab} is a Laurent polynomial in $q$ with integer coefficients. Also, when the gauge group is $U(N)$, monopole number sectors are labeled by an $N$-tuple of integers $m=(m_1,m_2,\ldots,m_N)$ and a given sector can only contribute to the index if $\sum m_i=0$.

\setlength\extrarowheight{6pt}
\begin{table}
 \begin{adjustwidth}{-1.7cm}{}
\footnotesize
    \begin{tabular}{ | x{0.6cm} | c | c | c | c | c | c |}
      \hline
       & $p=1$ & $p=2$ & $p=3$ & $p=4$ & $p=5$ & $p=6$ \\ \hhline{|=|=|=|=|=|=|=|}
      $U(2)$ &  $2(1-q)(1-q^{-1})$ & $4$ & $3$ & $3$ & $3$ & $3$\\[0.5em]  \hline
      $U(3)$ &  \begin{tabular}[c]{@{}l@{}} $6(1-q)^2(1-q^2)$\\ $(1-q^{-1})^2(1-q^{-2})$ \end{tabular} &  \begin{tabular}[c]{@{}l@{}}$28 - 6 q^{-2} - 8 q^{-1}$\\ $- 8 q - 6 q^2$ \end{tabular}& \begin{tabular}[c]{@{}l@{}}  $23 + 2 q^{-1} + 2q$ \end{tabular}& $16$ & $15$ & $15$ \\ \hline
      $U(4)$ &  \begin{tabular}[c]{@{}l@{}} $24(1-q)^3(1-q^2)^2$\\ $(1-q^3)(1-q^{-1})^3$\\ $(1-q^{-2})^2(1-q^{-3})$ \end{tabular}  & \begin{tabular}[c]{@{}l@{}} $504 +$\\ $84q^{-4} - 96q^{-3}$\\ $- 80q^{-2} - 160q^{-1}$\\ $- 160 q - 80 q^2$\\ $- 96 q^3+ 84 q^4$ \end{tabular}& \begin{tabular}[c]{@{}l@{}} $204 - 30q^{-3}$\\ $- 48q^{-2}- 24q^{-1}$\\ $- 24 q- 48 q^2$\\ $- 30 q^3$ \end{tabular} & \begin{tabular}[c]{@{}l@{}} $188 + 10q^{-2}$\\ $+ 24q^{-1}+ 24 q$\\ $+ 10 q^2$ \end{tabular} & \begin{tabular}[c]{@{}l@{}} $121 +$\\ $2q^{-1} + 2 q$ \end{tabular} & $108$ \\[0.6em] \hline 
      $U(5)$ &  \begin{tabular}[c]{@{}l@{}} $120(1-q)^4(1-q^2)^3$\\ $(1-q^3)^2 (1-q^4)$\\ $ (1-q^{-1})^4(1-q^{-2})^3$\\ $(1-q^{-3})^2(1-q^{-4})$ \end{tabular} & \begin{tabular}[c]{@{}l@{}}$12336+$\\ $120 q^{-10}+192 q^{-9}$\\ $-1080q^{-8}+48 q^{-7}$\\ $+120 q^{-6} +3792q^{-5}$\\ $-2016q^{-4}-1296q^{-3}$\\ $-3312q^{-2}-2736q^{-1}$\\ $-2736q -3312q^2$\\ $-1296q^3-2016q^4$\\ $+3792q^5+120q^6$\\ $+48q^7-1080q^8$\\ $+192q^9+120q^{10}$ \end{tabular} &  \begin{tabular}[c]{@{}l@{}} $3988+$\\ $180q^{-6}+388q^{-5}$\\ $-294q^{-4}-932q^{-3}$\\ $-584q^{-2}-752q^{-1}$\\ $-752q-584q^2$\\ $-932q^3-294q^4$\\ $+388q^5+180q^6$ \end{tabular} & \begin{tabular}[c]{@{}l@{}} $2144-$\\ $240q^{-4} -320q^{-3}$\\ $-320q^{-2}-192q^{-1}$\\ $-192q-320q^2$\\ $-320q^3-240q^4$ \end{tabular}  & \begin{tabular}[c]{@{}l@{}} $1897+$\\ $70q^{-3}+192q^{-2}$\\ $352q^{-1}+352q$\\ $+192q^2+70q^3$ \end{tabular}  & \begin{tabular}[c]{@{}l@{}} $1188+$\\ $14q^{-2}+40q^{-1}$\\ $40q+14q^2$ \end{tabular}   \\[0.5em]  \hline
    \end{tabular}
    \caption{The superconformal index of the ``Lens space theory'' $T[L(p,1),U(N)]$, which agrees with the partition function of $GL(N, \mathbb{C})$ Chern-Simons theory at level $k=0$ on Lens space $L(p,1)$.}    \label{IndexTab}
 \end{adjustwidth}
\end{table}   

From the table, one may be able to recognize the large $p$ behavior for $U(3)$ and $U(4)$ similar to \eqref{SU2Limit0} and \eqref{SU2Limit}. Indeed, it is a general feature of the index $\CI_{T[L(p,1);U(N)]}$ that fewer monopole number sectors contribute when $p$ increases. In order for a monopole number $m=(m_1,\ldots,m_N)$ to contribute, 
\be
|pm_i|\leq 2N-2
\ee
needs to be satisfied for all $m_i$. For large $p>2N-2$, $\CI$ only receives a contribution from the $m=0$ sector and becomes a constant: 
\be\label{Limit}
\CI(U(N),p>2N-2)=\CI_{m=(0,0,0,\ldots,0)}=(2N-1)!!\,.
\ee
For $p=2N-2$, the index receives contributions from two sectors\footnote{Here, double factorial of a negative number is taken to be 1.}:
\be
\CI(U(N),p=2N-2)=\CI_{m=(0,0,0,\ldots,0)}+\CI_{m=(1,0,\ldots,0,-1)}=\left[(2N-1)!!+(2N-5)!!\right].
\ee
While the $\ln q$ factors (that we have omitted) are artifacts of our scheme of removing zeros in $\CI$, the constant coefficient $(2N-1)!!$ in \eqref{Limit} is counting BPS states. Then one can ask a series of questions: 1) What are the states or local operators that are being counted? 2) Why is the number of such operators independent of $p$ when $p$ is large?    

Partition functions $Z_{\text{CS}}$ of the complex Chern-Simons theory on Lens spaces can also be computed systematically. Please see appendix \ref{Sec:CCS} for details of the method we use. For $k=0$, $G_\C=GL(N,\C)$,  the partition functions on $L(p,1)$ only depend on $q=e^{4\pi i/\tau}$ as $\bar{q}=e^{4\pi i/\bar{\tau}}=q^{-1}$.  After dropping a $(\ln q)^N$ factor as in the index case, it is again a polynomial. We have computed this partition function up to $N=5$ and $p=6$ and found a perfect agreement with the index in table \ref{IndexTab}. 

From the point of view of the complex Chern-Simons theory, this large $p$ behavior \eqref{Limit} seems to be even more surprising --- it predicts that the partition functions of the complex Chern-Simons theory on $L(p,1)$ at level $k=0$ are constant when $p$ is greater than twice the rank of the gauge group. One can then ask 1) why is this happening? And 2) what is the geometric meaning of this $(2N-1)!!$ constant?

\subsection{$T[L(p,1)]$ on $S^3_b$}

In previous sections, we have seen that the superconformal index of $T[L(p,1)]$ agrees completely with the partition function of the complex Chern-Simons theory at level $k=0$ given by \eqref{FullTree1} with trivial relative phases $C_\alpha=0$:
\be\label{FullTree2}
Z(G_\C;\tau,\bar{\tau})=\sum_{\alpha\in \CM} Z'_{\alpha}\left(G;\frac{\tau}{2}\right)Z'_{\alpha}\left(G;\frac{\bar{\tau}}{2}\right),
\ee
for $G=U(N)$. But for more general $k$, one can no longer expect this to be true. We will now consider the $S^3_b$ partition function of $T[L(p,1)]$, which will give the partition function of the complex Chern-Simons theory at level \cite{Cordova:2013cea}
\be
(k,\sigma)=\left(1,\frac{1-b^2}{1+b^2}\right).
\ee
And we will examine for which choices of $N$ and $p$ that setting all phases $C_\alpha=0$ becomes a mistake, by comparing the $S^3_b$ partition function of $T[L(p,1)]$ to the ``naive'' partition function \eqref{FullTree2} of the complex Chern-Simons theory at level $k=1$ on $L(p,1)$.

There are two kinds of squashed three-spheres breaking the $SO(4)$ isometry of the round $S^3$: the first one preserves $SU(2) \times U(1)$ isometry while the second one preserves $U(1) \times U(1)$ \cite{Hama:2011ea}. However, despite the geometry being different, the partition functions of 3d $\CN=2$ theories that one gets are the same \cite{Hama:2011ea, Imamura:2013nra, Martelli:2011fu, Martelli:2011fw}. In fact, as was shown in \cite{Alday:2013lba,Closset:2013vra}, three-sphere partition functions of $\CN=2$ theories only admit a one-parameter deformation. We will choose the ``ellipsoid'' geometry with the metric
\begin{equation}
ds_3^2 = f(\theta)^2 d\theta^2 +\cos^2 \theta d\phi_1^2 + \frac{1}{b^4} \sin^2 \theta d\phi_2^2,
\end{equation}
where $f(\theta)$ is arbitrary and does not affect the partition function of the supersymmetric theory. 

Using localization, partition function of a $\CN=2$ gauge theory on such an ellipsoid can be written as an integral over the Cartan of the gauge group \cite{Hama:2011ea, Martelli:2011fu}. Consider an $\CN=2$ Chern-Simons-matter theory with gauge group being $U(N)$. A classical Chern-Simons term with level $k$ contributes 
\begin{equation}
Z_{\text{CS}} = \exp  \left( \frac{i}{b^2} \frac{k}{4 \pi} \sum_{i=1}^{N} \lambda_i^2 \right)
\end{equation}
to the integrand. The one-loop determinant of $U(N)$ vector multiplet, combined with the Vandermonde determinant, gives
\begin{equation}
Z_{\text{gauge}}=\prod_{i < j}^{N} \left(2 \sinh \frac{\lambda_i - \lambda_j}{2} \right) \left(2 \sinh \frac{\lambda_i - \lambda_j}{2 b^2} \right).
\end{equation}
A chiral multiplet in the representation $\mathfrak{R}$ gives a product of double sine functions:
\begin{equation}
Z_{\text{matter}} = \prod_{\rho \in \mathfrak{R}} s_b \left( \frac{i Q}{2} \left( 1- R \right) - \frac{\rho(\lambda)}{2 \pi b} \right),
\end{equation}
where $Q = b + 1/b$, $R$ is the R-charge of the multiplet and the double sine function is defined as 
\begin{equation}
s_b(x) = \prod_{p,q = 0}^{+\infty} \frac{pb + qb^{-1} + \frac{Q}{2} - i x}{pb^{-1}+q b + \frac{Q}{2} + ix}.
\end{equation}

Then we can express the $S^3_b$ partition function of $T[L(p,1)]$ using the UV description in \eqref{LensTheory} as
\begin{equation}
\begin{aligned}
Z(T[L(p,1),U(N)], b) = \frac{1}{N!} \int  & \prod_i^{N} \frac{d\lambda_i}{2\pi} \exp  \left( -\frac{i}{b^2} \frac{p}{4 \pi} \sum_{i=1}^{N} \lambda_i^2 \right)\\
\times & \prod_{i < j}^{N} \frac{4}{\pi^2} \left( \sinh \frac{\lambda_i - \lambda_j}{2} \right)^2 \left( \sinh \frac{\lambda_i - \lambda_j}{2 b^2} \right)^2,
\end{aligned}
\end{equation}
which is a Gaussian integral. We list our results in table \ref{S3b tab1} and \ref{S3b tab2}. A universal factor 
\be
\left(\frac{b}{i p }\right)^{N/2}\pi^{-N(N-1)}
\ee
is dropped in making these two tables.

\setlength\extrarowheight{6pt}
\begin{table}[htbp]
\begin{adjustwidth}{-0.6cm}{}
    \centering\tiny\setlength\tabcolsep{1pt}
        \begin{tabular}{| x{0.8cm} | c | c | c |  }
           \hline
            $p$ &$U(2)$ &$U(3)$ & $U(4)$ \\[0.5em]
           \hhline{|=|=|=|=|}
             $\ 1\ $ & \begin{tabular}[c]{@{}l@{}} $\ \ 2 e^{-2 i \pi  b^2-\frac{2 i \pi }{b^2}}$\\ $ \left(1-e^{\frac{2 i \pi }{b^2}}\right)\left(1-e^{2 i \pi  b^2}\right)$ \end{tabular} & \begin{tabular}[c]{@{}l@{}}$6 e^{-8 i \pi  b^2-\frac{8 i \pi }{b^2}} \left(1-e^{\frac{2 i \pi }{b^2}}\right)^3 \left(1+e^{\frac{2 i \pi }{b^2}}\right)$\\ $\left(1-e^{2 i \pi  b^2}\right)^3 \left(1+e^{2 i \pi  b^2}\right)$ \end{tabular} & \begin{tabular}[c]{@{}l@{}} $24 e^{-20 i \pi b^2-\frac{20 i \pi} {b^2}}  \left(1-e^{\frac{2 i \pi }{b^2}}\right)^6\left(1+e^{\frac{2 i \pi }{b^2}}\right)^2$\\ $\left(1+e^{\frac{2 i \pi }{b^2}}+e^{\frac{4 i \pi
   }{b^2}}\right)\left(1-e^{2 i \pi  b^2}\right)^6$\\ $\left(1+e^{2 i \pi  b^2}\right)^2 \left(1+e^{2 i
   \pi  b^2}+e^{4 i \pi  b^2}\right)$ \end{tabular}        \\ \hline
             $\ 2\ $ &\begin{tabular}[c]{@{}l@{}}$ 2-2 e^{-\frac{i \pi }{b^2}}-2 e^{-i \pi  b^2}$\\[0.5em] $+2 e^{-i \pi  b^2-\frac{i \pi }{b^2}}$ \end{tabular}&  \begin{tabular}[c]{@{}l@{}}$2 e^{-4 i \pi  (b^2+b^{-2})}$\\[0.5em] $ \left(1-e^{\frac{2 i \pi }{b^2}}\right) \left(1-e^{2 i\pi  b^2}\right)$\\[0.5em] $ \left(-6 e^{\frac{i \pi }{b^2}}+3 e^{\frac{2 i \pi }{b^2}}-6 e^{i \pi  b^2}+3 e^{2 i\pi  b^2}\right.$\\ $\left.-4 e^{ i \pi  (b^2+b^{-2})}+3 e^{ 2 i \pi  (b^2+b^{-2})}\right.$\\[0.5em] $\left. -6e^{i \pi  \left(b^2+2 b^{-2}\right)}-6 e^{i \pi \left(2  b^2+ b^{-2} \right)}+3 \right)$ \end{tabular}  & \begin{tabular}[c]{@{}l@{}}  $8 e^{-10 i \pi \left(b^2+b^{-2}\right)} \left(1-e^{\frac{2 i \pi }{b^2}}\right)^2 \left(1-e^{2 i b^2 \pi }\right)^2 $\\ $ \left(3-9 e^{\frac{i \pi }{b^2}}+9 e^{\frac{2 i \pi }{b^2}}-6 e^{\frac{3 i \pi }{b^2}}+9 e^{\frac{4 i \pi }{b^2}}-9 e^{\frac{5 i \pi }{b^2}} \right.$\\ $\left.+3 e^{\frac{6 i \pi}{b^2}}-9 e^{i b^2 \pi }+9 e^{2 i b^2 \pi }-6 e^{3 i b^2 \pi } \right.$\\ $\left.+9 e^{4 i b^2 \pi }-9 e^{5 i b^2 \pi}+3 e^{6 i b^2 \pi }-9 e^{i \pi \left(b^2+b^{-2}\right)} \right.$\\ $\left.+27 e^{2 i \pi \left(b^2+b^{-2}\right)}-4 e^{3 i \pi \left(b^2+b^{-2}\right) }+27 e^{4 i \pi \left(b^2+b^{-2}\right)
   } \right.$\\ $\left.-9 e^{ 5 i \pi \left(b^2+b^{-2}\right) }+3 e^{6 i \pi \left(b^2+b^{-2}\right)}-27 e^{i \pi \left(b^2+2b^{-2}\right) } \right.$\\ $\left.+27 e^{2 i \pi \left(b^2+2 b^{-2}\right)}-6e^{3 i \pi \left(b^2+2b^{-2}\right)}-6 e^{i \pi \left(b^2+3b^{-2}\right) } \right.$\\ $\left.+9 e^{2 i \pi \left(b^2+3b^{-2}\right)}-27 e^{i \pi \left(b^2+4b^{-2}\right) } -9 e^{i \pi \left(b^2+5 b^{-2} \right)} \right.$\\ $\left.-9 e^{i \pi \left(b^2+6b^{-2}\right)}-18 e^{i \pi \left(2b^2+3b^{-2}\right) }+9 e^{2 i \pi \left(2 b^2+3b^{-2}\right)} \right.$\\ $\left.-27 e^{i \pi \left(2b^2+5b^{-2}\right)}-18 e^{i \pi \left(3 b^2+2b^{-2}\right) }+9 e^{2 i \pi \left(3b^2+2b^{-2}\right)} \right.$\\ $\left.-18 e^{i \pi \left(3 b^2+4b^{-2}\right) }-6 e^{i \pi \left(3b^2+5b^{-2}\right) }-18 e^{i \pi \left(4 b^2+3b^{-2}\right)} \right.$\\ $\left.-27 e^{i \pi \left(4b^2+5b^{-2}\right) }-27 e^{i \pi \left(5 b^2+2b^{-2}\right)}-6 e^{i \pi \left(5b^2+3b^{-2}\right)} \right.$\\ $\left.-27 e^{i \pi \left(5 b^2+4b^{-2}\right) }-9 e^{i \pi \left(5b^2+6b^{-2}\right)}-9 e^{i \pi \left(6 b^2+5b^{-2}\right) } \right.$\\ $\left.-27 e^{i \pi \left(2 b^2+ b^{-2} \right)}+27 e^{2 i \pi \left(2b^2+b^{-2} \right)} \right.$\\ $\left.-6 e^{3 i \pi \left(2 b^2+b^{-2}\right)}-6 e^{i \pi \left(3b^2+b^{-2}\right)}+9 e^{2 i \pi \left(3 b^2+b^{-2} \right)} \right.$\\ $\left.-27 e^{i \pi \left(4 b^2+b^{-2} \right)}-9 e^{i \pi \left(5 b^2+ b^{-2} \right)}-9 e^{i \pi \left(6 b^2+b^{-2} \right)}\right)$ \end{tabular}   \\ \hline
             $\ 3\ $ & \begin{tabular}[c]{@{}l@{}} $2-2 e^{-\frac{2 i \pi }{3 b^2}}-2 e^{-\frac{2}{3} i \pi  b^2}$\\[0.5em] $-e^{-\frac{2 i \pi}{3} (b^2+b^{-2})}$ \end{tabular}    & \begin{tabular}[c]{@{}l@{}} $-3 e^{-\frac{8 i \pi}{3}  \left(b^2+b^{-2}\right)}\times $\\ $ \left(4 e^{\frac{2 i \pi }{3 b^2}}+2 e^{\frac{2 i \pi}{b^2}}+2 e^{\frac{8 i \pi }{3 b^2}} \right.$\\[0.5em] $\left. +4 e^{\frac{2}{3} i \pi  b^2}+2 e^{2 i \pi  b^2}+2 e^{\frac{8}{3} i \pi  b^2}\right. $\\[0.5em] $ \left. -8 e^{\frac{2 i \pi}{3}  \left(b^2+b^{-2}\right)}+4 e^{2 i \pi \left(b^2+b^{-2}\right)}\right.$\\[0.5em] $\left.-2 e^{\frac{8 i \pi}{3}  \left(b^2+b^{-2}\right)}+8 e^{\frac{2 i \pi}{3} \left(b^2+3b^{-2}\right)}\right.$\\[0.5em] $\left.-4 e^{\frac{2 i \pi}{3}  \left(b^2+4b^{-2}\right)}\right.$\\[0.5em] $\left.+4 e^{\frac{2 i \pi}{3} \left(3 b^2+4b^{-2}\right)}+4 e^{\frac{2 i \pi}{3}  \left(4 b^2+3b^{-2}\right)}\right.$\\[0.5em] $\left. +8 e^{\frac{2 \pi i}{3} \left(3  b^2+b^{-2}\right)}-4 e^{\frac{2\pi i}{3} \left(4 b^2+\pi b^{-2} \right)}+1\right)$ \end{tabular}     &  \begin{tabular}[c]{@{}l@{}}  $-6 e^{-\frac{20 i \pi}{3} \left(b^2+b^{-2}\right)} \left(1-e^{\frac{2 i \pi }{b^2}}\right)
   \left(1-e^{2 i b^2 \pi }\right)$\\ $\left(1+6 e^{\frac{2 i \pi }{3 b^2}}+5 e^{\frac{2 i \pi }{b^2}}+8
   e^{\frac{8 i \pi }{3 b^2}}+3 e^{\frac{4 i \pi }{b^2}}+4 e^{\frac{14 i \pi }{3 b^2}}\right.$\\ $\left.+6 e^{\frac{2}{3}
   i b^2 \pi }+5 e^{2 i b^2 \pi }+8 e^{\frac{8}{3} i b^2 \pi }+3 e^{4 i b^2 \pi }\right.$\\ $\left.+4 e^{\frac{14}{3} i
   b^2 \pi }-18 e^{\frac{2 i \pi}{3} \left(b^2+b^{-2}\right)}-2 e^{\frac{4 i \pi}{3} \left(b^2+b^{-2}\right)}\right.$\\ $\left.+25 e^{2 i pi \left(b^2+b^{-2}\right)}-28 e^{\frac{8 i \pi}{3} \left(b^2+b^{-2}\right) }-2 e^{\frac{10 i \pi}{3} \left(b^2+b^{-2} \right) }\right.$\\ $\left.+9 e^{4 i \pi \left(b^2+b^{-2}\right) }-4 e^{\frac{14 i \pi}{3} \left(b^2+b^{-2}\right) }-4 e^{\frac{4 i \pi}{3} \left(b^2+2b^{-2} \right)}\right.$\\ $\left.+15 e^{2 i \pi \left(b^2+2b^{-2} \right) }+30 e^{\frac{2 i \pi}{3} \left(b^2+3b^{-2} \right)}-24 e^{\frac{2 i \pi}{3} \left(b^2+4b^{-2} \right) }\right.$\\ $\left.+18 e^{\frac{2 i \pi}{3} \left(b^2+6b^{-2} \right)}-12 e^{\frac{2 i \pi}{3} \left(b^2+7b^{-2} \right)}\right.$\\ $\left.+24 e^{\frac{4 i \pi}{3} \left(2 b^2+3b^{-2}\right)}+2 e^{\frac{2 i \pi}{3} \left(2 b^2+5b^{-2} \right)}+4 e^{\frac{2 i \pi}{3} \left(2 b^2+7b^{-2} \right) }\right.$\\ $\left.+24 e^{\frac{4 i \pi}{3} \left(3 b^2+2b^{-2} \right)}+40 e^{\frac{2 i \pi}{3} \left(3 b^2+4b^{-2} \right)}+20 e^{\frac{2 i \pi}{3} \left(3 b^2+7b^{-2} \right)}\right.$\\ $\left.+40 e^{\frac{2 i \pi}{3}\left(4 b^2+3b^{-2} \right)}+4 e^{\frac{2 i \pi}{3}  \left(4 b^2+5b^{-2}\right)}-20 e^{\frac{2 i \pi}{3} \left(4 b^2+7b^{-2} \right) }\right.$\\ $\left.+2 e^{\frac{2 i \pi}{3}  \left(5 b^2+2b^{-2} \right)}+4 e^{\frac{2 i \pi}{3} \left(5 b^2+4b^{-2} \right) }-4 e^{\frac{2 i \pi}{3} \left(5 b^2+7b^{-2}\right) }\right.$\\ $\left.+12 e^{\frac{2 i \pi}{3} \left(6 b^2+7b^{-2} \right) }+4 e^{\frac{2 i \pi}{3} \left(7 b^2+2b^{-2} \right)}+20 e^{\frac{2 i \pi}{3} \left(7 b^2+3b^{-2}\right) }\right.$\\ $\left.-20 e^{\frac{2 i \pi}{3}  \left(7 b^2+4b^{-2}\right)}-4 e^{\frac{2 i \pi}{3} \left(7 b^2+5b^{-2}\right) }+12 e^{\frac{2 i \pi}{3} \left(7 b^2+6b^{-2}\right) }\right.$\\ $\left.-4 e^{\frac{4 i \pi}{3} \left(2b^2+b^{-2}\right)}+15 e^{2 i \pi \left(2 b^2+b^{-2} \right)}+30 e^{\frac{2 i \pi}{3} \left(3 b^2+b^{-2} \right)}\right.$\\ $\left.-24 e^{\frac{2 i \pi}{3} \left(4 b^2+b^{-2} \right)}+18 e^{\frac{2 i \pi}{3}\left(6 b^2+b^{-2} \right)}-12 e^{\frac{2 i \pi}{3} \left(7 b^2+b^{-2} \right)}\right)$ \end{tabular}       \\ \hline

        \end{tabular}
        \caption{The $S^3_b$ partition function of $T[L(p,1),U(N)]$. In this table $p$ ranges from $1$ to $3$.}
               \label{S3b tab1}
   \end{adjustwidth}
\end{table}

\setlength\extrarowheight{12pt}
\begin{table}[ht]
\begin{adjustwidth}{-0.6cm}{}
    \centering\footnotesize\setlength\tabcolsep{3pt}
        \hspace*{-1cm}\begin{tabular}{| x{0.6cm}  | c | c |  }
           \hline
             $p$ & $U(2)$ & $U(3)$  \\[0.6em]
           \hhline{|=|=|=|}
             $\ 4\ $ &  $2-2 e^{-\frac{i \pi }{2 b^2}}-2 e^{-\frac{1}{2} i \pi  b^2}-2e^{-\frac{i \pi}{2}  \left(b^2+b^{-2}\right)}$ & \begin{tabular}[c]{@{}l@{}}$-2 e^{-2 i \pi  \left(b^2+b^{-2}\right)}\times $\\ $ \left(-3-2 e^{\frac{i \pi }{2 b^2}}+2 e^{\frac{3 i \pi }{2b^2}}+3 e^{\frac{2 i \pi }{b^2}}-2 e^{\frac{1}{2} i \pi  b^2}+2 e^{\frac{3}{2} i \pi  b^2}+3 e^{2 i\pi  b^2}+4 e^{\frac{i \pi}{2}  \left(b^2+b^{-2}\right)}\right.$\\ $\left.+4 e^{\frac{3 i \pi}{2}\left(b^2+b^{-2}\right)}-3 e^{2 i \pi  \left(b^2+b^{-2}\right)}+4 e^{\frac{i \pi}{2}  \left(b^2+3b^{-2}\right)}-6e^{\frac{i \pi}{2}  \left(b^2+4b^{-2}\right)}\right.$\\ $\left.+6 e^{\frac{i \pi}{2}  \left(3 b^2+4b^{-2}\right)}+6e^{\frac{i \pi}{2}  \left(4 b^2+3b^{-2}\right)}+4 e^{\frac{i \pi}{2}\left(3 b^2+b^{-2} \right)}-6e^{\frac{i \pi}{2} \left(4b^2+b^{-2} \right)}\right)$ \end{tabular}  \\[1.5em] \hline
             $\ 5\ $ &\begin{tabular}[c]{@{}l@{}}$2-2 e^{-\frac{2 i \pi }{5 b^2}}-2 e^{-\frac{2}{5} i \pi  b^2}+2\cos \frac{4\pi}{5} e^{-\frac{2 i \pi}{5}  \left(b^2+b^{-2}\right)}$ \end{tabular} &\begin{tabular}[c]{@{}l@{}} $6-12 e^{-\frac{2 i \pi }{5 b^2}}+12 e^{-\frac{6 i \pi }{5 b^2}}-6 e^{-\frac{8 i \pi }{5 b^2}}-12e^{-\frac{2}{5} i \pi  b^2}$\\ $+12 e^{-\frac{6}{5} i \pi  b^2}-6 e^{-\frac{8}{5} i \pi  b^2}+4\left(\cos \frac{8\pi}{5} +e^{\frac{4 i \pi}{5}} \right)e^{-\frac{2 i \pi}{5} (4b^2+b^{-2})}$\\ $ 4\left(\cos \frac{8\pi}{5} +2 \cos \frac{4\pi}{5} \right) e^{-\frac{2i\pi}{5} (b^2+4b^{-2})} + 8\left( \cos \frac{4 \pi}{5} + 2 \cos \frac{2\pi}{5} \right) e^{-\frac{2 i \pi}{5} \pbra{b^2+b^{-2}}}$\\ $+ 8\left(\cos \frac{12\pi}{5} + 2 \cos \frac{6 \pi}{5} \right) e^{-\frac{6 i \pi}{5}\left(b^2+b^{-2}\right)}+2\left(\cos \frac{16 \pi}{5} + 2\cos \frac{8\pi}{5} \right)$\\ $\times e^{-\frac{8 i \pi}{5}\left(b^2+b^{-2}\right)} -8 e^{-\frac{2 i \pi}{5}  \left(b^2+3b^{-2}\right)}-8 e^{-\frac{2 i \pi}{5}  \left(b^2-3+3b^{-2}\right)}-8 e^{-\frac{2 i \pi}{5} \left(b^2+3+3b^{-2}\right)}$\\ $-8 e^{-\frac{2 i \pi}{5}\left(3b^2+b^{-2}\right)}-4 e^{-\frac{2 i \pi}{5}\left(3 b^2+4b^{-2}\right)}-4 e^{-\frac{2 i \pi}{5} \left(3b^2-6+4b^{-2}\right)}$\\ $-8 e^{-\frac{2 i \pi}{5}  \left(3 b^2-3+b^{-2}\right)}-8 e^{-\frac{2 i\pi}{5} \left(3 b^2+3+b^{-2}\right)}-4 e^{-\frac{2 i \pi}{5}  \left(3 b^2+6+4b^{-2}\right)}$\\ $-4e^{-\frac{2 i \pi}{5} \left(4b^2+3b^{-2}\right)}-4 e^{-\frac{2 i \pi}{5}  \left(4 b^2-6+3b^{-2}\right)}-4 e^{-\frac{2 i \pi}{5} \left(4 b^2+6+3b^{-2}\right)}$ \end{tabular}  \\ \hline
             $\ 6\ $ &\begin{tabular}[c]{@{}l@{}} $2-2 e^{-\frac{ i \pi }{3 b^2}}-2 e^{-\frac{1}{3} i \pi  b^2}+e^{-\frac{ i \pi}{3} (b^2+b^{-2})}$ \end{tabular} & \begin{tabular}[c]{@{}l@{}} $ e^{-\frac{4 i \pi}{3} \left(b^2+b^{-2}\right)}\times $\\ $\left(-12 e^{\frac{i \pi }{3 b^2}}-6 e^{\frac{i \pi}{b^2}}-6 e^{\frac{4 i \pi }{3 b^2}}-12 e^{\frac{1}{3} i \pi  b^2}-6 e^{i \pi  b^2}-6 e^{\frac{4}{3}i \pi  b^2}-8 e^{\frac{i \pi}{3}  \left(b^2+b^{-2}\right)}\right.$\\ $\left.+4 e^{i \pi \left(b^2+b^{-2}\right)}+6 e^{\frac{4 i \pi}{3}  \left(b^2+b^{-2}\right)}+8 e^{\frac{i \pi}{3} \left(b^2+3b^{-2}\right)}+12 e^{\frac{i \pi}{3}  \left(b^2+4b^{-2}\right)}\right.$\\ $\left.-12 e^{\frac{i \pi}{3}  \left(3b^2+4b^{-2}\right)}-12 e^{\frac{i \pi}{3}\left(4 b^2+3b^{-2}\right)}+8 e^{\frac{i \pi}{3} \left(3 b^2+b^{-2} \right)}+12 e^{\frac{i\pi }{3} \left(4  b^2+b^{-2} \right)}-3\right)$ \end{tabular}  \\ \hline

        \end{tabular}\hspace*{-1cm}
        \caption{The $S^3_b$ partition function of $T[L(p,1),U(N)]$. This table, with $p$ ranging from $4$ to $6$, is the continuation of the previous table \ref{S3b tab1}. Due to the limitation of space, only partition functions for $U(2)$ and $U(3)$ are given.}
               \label{S3b tab2}
  \end{adjustwidth}
\end{table}

If one compares results in table \ref{S3b tab1} and \ref{S3b tab2} with partition functions of complex Chern-Simons theory naively computed using \eqref{FullTree}, one will find a perfect agreement for $p=1$ once the phase factor
\be
\exp\left[ \frac{\pi i (c_L-c_R)}{12}\cdot (3-p) \right]
\ee
from the change of framing is added\footnote{The complex Chern-Simons theory obtained from the 3d-3d correspondence is naturally in ``Seifert framing'', as the $T[L(p,1)]$ we used is obtained by reducing M5-brane on the Seifeit $S^1$ fiber of $L(p,1)$ in \cite{equivariant}. However, the computation in appendix \ref{Sec:CCS} is in ``canonical framing'' and differs from Seifert framing by $(3-p)$ units \cite{Beasley:2005vf}. }. This agreement is not unexpected because for $p=1$, $\CM_{\text{flat}}$ consists of just a single point and there are no such things as relative phases between contributions from different flat connections. Even for $p=2$, the naive way \eqref{FullTree} of computing partition function of complex Chern-Simons theory seems to be still valid modulo an overall factor. However, starting from $p=3$, the two sides start to differ significantly. See table \ref{S3b tab0} for a comparison between the $S^3_b$ partition function of $T[L(p,1)]$ and the ``naive'' partition function of the complex Chern-Simons theory on $L(p,1)$ for $G=U(2)$.

\setlength\extrarowheight{8pt}
\begin{table}[htbp]
 \begin{adjustwidth}{-1.0cm}{}
    \centering\footnotesize\setlength\tabcolsep{1pt}
        \begin{tabular}{| x{0.5cm} | c | c |   }
           \hline
            $p$ & $S^3_b$ partition function of $T[L(p,1); U(2)]$ & ``naive'' partition function of $GL(2,\C)$ Chern-Simons theory \\[0.5em]
           \hhline{|=|=|=|}
             $\ 1\ $ & \begin{tabular}[c]{@{}l@{}}$ 2-2 q^{-1}-2 \tq^{-1}+2 \pbra{q\tq}^{-1}$ \end{tabular}& \begin{tabular}[c]{@{}l@{}} $2-2 q^{-1}-2 \tq^{-1}+2 \pbra{q\tq}^{-1}$ \end{tabular}     \\[0.7em] \hline
             $\ 2\ $ &\begin{tabular}[c]{@{}l@{}}$ 2+2 q^{-\frac{1}{2}}+2 \tq^{-\frac{1}{2}}+2 \pbra{q\tq}^{-\frac{1}{2}}$ \end{tabular}&  \begin{tabular}[c]{@{}l@{}} $2i(2+2 q^{-\frac{1}{2}}+2 \tq^{-\frac{1}{2}}+2 \pbra{q\tq}^{-\frac{1}{2}})$\end{tabular}    \\[0.7em] \hline
             $\ 3\ $ & \begin{tabular}[c]{@{}l@{}} $ 2+\pbra{1-\sqrt{3} i} q^{-\frac{1}{3}}+\pbra{1-\sqrt{3} i} \tq^{-\frac{1}{3}}+\frac{1}{2}\pbra{1+\sqrt{3}i} \pbra{q\tq}^{-\frac{1}{3}}$ \end{tabular}    & \begin{tabular}[c]{@{}l@{}} $2 +\left(1-3\sqrt{3}i\right)\tq^{\frac{1}{3}}+\left(1-3\sqrt{3}i\right) q^{\frac{1}{3}}+\frac{1}{2} \left(1+3 \sqrt{3} i \right) \pbra{q \tq}^{\frac{1}{3}} $ \end{tabular}     \\[0.7em] \hline
             $\ 4\ $ &  $ 2-2iq^{-\frac{1}{4}}-2i \tq^{-\frac{1}{4}}+2 \pbra{q\tq}^{-\frac{1}{4}}$& $8 i \pbra{q \tq}^{\frac{1}{2}} \pbra{1+ i q^{\frac{1}{4}}+i\tq^{\frac{1}{4}}+ (q\tq)^{\frac{1}{4}}} $  \\[0.7em] \hline
             $\ 5\ $ &\begin{tabular}[c]{@{}l@{}}$2-2e^{\frac{2\pi i}{5}} q^{-\frac{1}{5}}-2 e^{\frac{2\pi i}{5}} \tq^{-\frac{1}{5}}+2\cos \frac{4\pi}{5}e^{\frac{4 \pi i}{5}} \pbra{q\tq}^{-\frac{1}{5}} $ \end{tabular} & \begin{tabular}[c]{@{}l@{}} $q \tq \left(2-2\left(e^{\frac{3 \pi i}{5}} + 2e^{\frac{4 \pi i}{5}}\right) \tq^{\frac{1}{5}}-2 \pbra{e^{\frac{3 \pi i}{5}}+2e^{\frac{4 \pi i}{5}}}q^{\frac{1}{5}}\right.$\\ $\left. +\left( 1+ 2e^{\frac{\pi i}{5}}+3e^{\frac{2 \pi i}{5}}-4e^{\frac{3 \pi i}{5}}-4e^{\frac{4\pi i}{5}} \right) (q \tq)^{\frac{1}{5}}\right) $\end{tabular} \\ \hline
             $\ 6\ $ &\begin{tabular}[c]{@{}l@{}} $ 2- \pbra{1+\sqrt{3} i}q^{-\frac{1}{6}}-\pbra{1+\sqrt{3} i} \tq^{-\frac{1}{6}}-\frac{1}{2}\pbra{1-\sqrt{3} i} \pbra{q\tq}^{-\frac{1}{6}}$ \end{tabular} & $6 i \pbra{q \tq}^{\frac{3}{2}} \pbra{2+(-1+i\sqrt{3})q^{\frac{1}{6}}+(-1+i\sqrt{3})\tq^{\frac{1}{6}}+ \frac{1}{2} \pbra{1+i\sqrt{3}}(q \tq)^{\frac{1}{6}}  }$ \\[0.7em] \hline

        \end{tabular}
        \caption{The comparison between the $S^3_b$ partition function of $T[L(p,1),U(2)]$ and the ``naive'' partition function of the $GL(2, \mathbb{C})$ Chern-Simons theory, obtained by putting together two copies of the $U(2)$ Chern-Simons theory using \eqref{FullTree2}, on Lens space $L(p,1)$ in ``Seifert framing.'' Notice that when $p$ increases, the difference between the two columns becomes larger and larger.}
               \label{S3b tab0}
  \end{adjustwidth}
\end{table}

\clearpage

\appendix
\section{Complex Chern-Simons theory on Lens spaces}\label{Sec:CCS}

Lens space $L(p,q)$ can be obtained by gluing two solid tori $S^1\times D^2$ along their boundary $T^2$'s using an element in $\mathrm{MCG}(T^2)=SL(2,\Z)$:
\begin{equation}
\left( \vspace{-1.0cm} \begin{array}{cc} \matminus q\  &\ *\\p\ &\ * \vspace{5pt} \end{array} \right)\left( \begin{array}{c}m\\l \vspace{5pt} \end{array}\right)=\left(\begin{array}{c}m'\\l' \vspace{5pt} \end{array}\right).
\label{MCG}
\end{equation}
Here $(m,l)$ and $(m',l')$ are meridian and longitude circles of the two copies of $T^2=\partial (S^1\times D^2)$. So the meridian $m'$ of one torus is mapped to $-qm+pl$ of the other torus. As for $l$, we do not need to track what it is mapped into as the choice only affects the framing of $L(p,q)$. A canonical choice of an $SL(2,\Z)$ element in \eqref{MCG} is given by
\be
ST^{c_1}ST^{c_2}S \ldots T^{c_n}S,
\ee
where $(c_1,c_2,\ldots,c_n)$ are coefficients in continued fraction expansion of $p/q$. For $q=1$, the element that gives $L(p,1)$ is
\be
ST^{p}S.
\ee

As $SL(2,\Z)$ naturally acts on the Hilbert space $\CH^{\text{CS}}(T^2;G)$ of the Chern-Simons theory on the two-torus, one has
\be\label{LPQ}
Z_{\text{CS}}(L(p,q);G)=\langle 0|\CS\CT^{c_1}\CS\CT^{c_2}\CS \ldots \CT^{c_n}\CS |0\rangle.
\ee
Here $|0\rangle \in \CH$ is the state associated to the solid torus while $\CS$ and $\CT$ give the action of $S,T\in SL(2,\Z)$ on $\CH$. When $G$ is compact, $\CS$ and $\CT$ are known from the study of the 2D WZW model and affine Lie algebra \cite{Kač1984125} and can be directly used to evaluate \eqref{LPQ}. Partition functions of Chern-Simons theory on Lens spaces were first obtained precisely in this manner in \cite{jeffrey1992} for $SU(2)$ and in \cite{Marino:2002fk, RT-invariant} for higher rank gauge groups. Define $\hat k = k + \check h$, then the partition function of the $G$ Chern-Simons theory on $L(p,q)$ is given by
\begin{equation}\label{CSFull}
\begin{aligned}
Z(L(p,q), \hat k) = & \frac{1}{(\hat k |p|)^{N/2}} \exp \left(\frac{i \pi}{\hat k} s(q,p) |\rho|^2 \right) \\
&\times \sum_{w \in W} \det (w) \exp \left(-\frac{2\pi i}{p \hat k}\langle \rho, w(\rho) \rangle \right)\\
& \times \sum_{m \in Y^{\vee}/pY^{\vee}} \exp \left(i \pi \frac{q}{p} \hat k |m|^2 \right) \exp \left(2 \pi i \frac{1}{p} \langle m, q \rho - w(\rho) \rangle \right).
\end{aligned}
\end{equation}
Here $s(q,p)$ is the Dedekind sum:
\begin{equation}
s(q,p) = \frac{1}{4 p} \sum_{n=1}^{p-1} \cot \left( \frac{\pi n}{p} \right) \cot \left( \frac{\pi q n}{p}\right) , 
\end{equation}
$\rho$ the Weyl vector of the Lie algebra $\mathfrak{g}$, $W$ the Weyl group, $Y^{\vee}$ the coroot lattice, $N$ the rank of the gauge group, and the inner product, $\langle\cdot,\cdot\rangle$, is taken with respect to the standard Killing form of $\mathfrak{g}$. 

Now we start computing the partition function of complex Chern-Simons theory using \eqref{FullTree1} for $G_\C=GL(N,\C)$. The first step is to separate \eqref{CSFull} into contributions from different flat connections. As discussed in section \ref{SSec:Space}, the moduli space $\CM_{\text{flat}}$ of $U(N)$ flat connections of $L(p,q)$ --- whose foundamental group is $\Z_p$ --- consists of discrete points. Each point can be labelled by $(a_1,a_2,\ldots,a_N)$, where the $a_j$'s are the $p$-th roots of unity. For convenience we use a different set of labels, $\alpha=(\alpha_1,\alpha_2,\ldots,\alpha_N)\in\frak{g}^*$, with the $\alpha_j$'s being integers between 0 and $p-1$ that satisfy
\be
e^{2\pi i \alpha_j/p}=a_j.
\ee
Then \eqref{CSFull} can be rewritten as \cite{Gang:2009wy}: 
\begin{equation}\label{CSSplit}
\begin{aligned}
Z(L(p,q),\hat k) & = \frac{1}{N!}\sum_{\alpha} Z_\alpha(L(p,q),\hat k),\\[0.5em]
Z_\alpha(L(p,q),\hat k) & = \frac{1}{(\hat k |p|)^{l/2}}  \exp \left(\frac{i \pi}{\hat k} N (N^2-1) s(q,p) \right) \exp \left(i \pi \frac{q}{p} \hat k |\alpha|^2 \right)\\
& \sum_{w,\tilde w \in S_N} \det (w) \exp \left(-\frac{2\pi i}{p \hat k}\langle \rho, w(\rho) \rangle \right) \exp \left(2 \pi i \frac{1}{p} \langle {\tilde w}(\alpha), q \rho - w(\rho) \rangle \right).
\end{aligned}
\end{equation}
The set $\{\alpha\}$ is redundant for labelling flat connections in $\CM_{\text{flat}}$ because the Weyl group $\CW=S_N\subset U(N)$ acts on $\{\alpha\}$ by permuting the $\alpha_j$'s. We will use $\tilde{\alpha}$ to denote equivalence classes of $\alpha$ under Weyl group action and each $\tilde{\alpha}$ corresponds to one flat connection modulo gauge transformations. A canonical representative of $\tilde{\alpha}$ is given by $(\alpha_1,\alpha_2,\ldots,\alpha_N)$ with $\alpha_1\geq\alpha_2\geq\ldots\geq\alpha_N$. Using $\tilde{\alpha}$, \eqref{CSFull} can be written as 
\be
Z(L(p,q),{\hat k})  = \sum_{\tilde{\alpha}}\frac{1}{|\CW_{\tilde{\alpha}}|} Z_{\tilde{\alpha}}(L(p,q),\hat k),
\ee 
where $\CW_{\tilde{\alpha}}\subset\CW$ is the stabilizer subgroup of $\tilde{\alpha}\in\frak{g}^*$.

Using the naive way \eqref{FullTree} of computing the partition function of complex Chern-Simons theory when $\CM_{\text{flat}}$ is zero-dimensional, one has 
\begin{equation}
Z(G_{\mathbb{C}}; \tau,\bar{\tau}) = \frac{1}{N!}\sum_{\alpha} Z_{\alpha} \left(G; \frac{\tau}{2}-\ch \right) Z_{\alpha} \left(G; \frac{\bar \tau}{2}-\ch \right).
\label{CS block}
\end{equation}
Notice that using $\tilde{\alpha}$ labels, this is
\be\label{CCSSplit}
Z(G_{\mathbb{C}}; \tau,\bar{\tau}) = \sum_{\tilde{\alpha}} \frac{1}{|\CW_{\tilde{\alpha}}|}Z_{\tilde{\alpha}} \left(G; \frac{\tau}{2}-\ch \right) Z_{\tilde{\alpha}} \left(G; \frac{\bar \tau}{2}-\ch \right),
\ee
and the $\frac{1}{|\CW_{\tilde{\alpha}}|}$ factor should not be squared. This is because $G_\C$ and $G$ have the same Weyl group $\CW$ and in complex Chern-Simons theory $\CW$ acts simultaneously on $\CA$ and $\bar{\CA}$. 

\eqref{CCSSplit}, together with \eqref{CSSplit}, is the equation we use to compute the partition function of the complex Chern-Simons theory. In the making of the table \ref{IndexTab}, we have dropped a universal factor 
\be
\left(\frac{4}{\tau\bar{\tau}}\right)^{N/2} \propto (\ln q)^N.
\ee
This matches the factor that is also omitted on the supersymmetric index side.

\acknowledgments{We are deeply indebted to Sergei Gukov for his valuable suggestions and constant encouragement at various stages of this work, and to Ingmar Saberi for proofreading our manuscript and giving very helpful comments. We also wish to thank Murat Kolo\u{g}lu, Petr Kravchuk, Pavel Putrov, Kung-Yi Su and Wenbin Yan for stimulating discussions. This work is funded by the DOE Grant DE-SC0011632 and the Walter Burke Institute for Theoretical Physics.}


\newpage

\bibliographystyle{JHEP_TD}
\bibliography{draft}

\end{document}